\documentclass[a4paper,12pt]{article}

\usepackage[font={small}]{caption}
\usepackage{amsmath}
\usepackage{color,graphicx}
\usepackage[top=2.5cm,right=2cm,bottom=2.5cm,left=2cm]{geometry}
\usepackage[colorlinks,linkcolor=blue,citecolor=blue,pagebackref=false]{hyperref}
\usepackage{cite}
\usepackage[parfill]{parskip}
\usepackage{multirow}
\usepackage{subcaption}

\begin{document}

\begin{center}

\textbf{Synaptic delays modulate population phase and amplitude\\ responses in oscillatory excitatory-inhibitory networks}

\vspace{0.5 cm}

Parsa Shahab Rad\textsuperscript{1},
Mojtaba Madadi Asl\textsuperscript{2,3,*}, and 
Alireza Valizadeh\textsuperscript{1,4,*} \let\thefootnote\relax\footnotetext{* Corresponding authors. Email addresses: m.madadi@ipm.ir, and valizadeh@gmail.com.}
\\
\bigskip
\footnotesize{
\textsuperscript{1}Department of Physics, Institute for Advanced Studies in Basic Sciences (IASBS), Zanjan, Iran\\
\textsuperscript{2}School of Biological Sciences, Institute for Research in Fundamental Sciences (IPM), Tehran, Iran\\
\textsuperscript{3}Pasargad Institute for Advanced Innovative Solutions (PIAIS), Tehran, Iran\\
\textsuperscript{4}The Zapata-Briceño Institute of Neuroscience, Madrid, Spain\\
}

\end{center}


\vspace{0.4cm}

\begin{center}


{\small

\begin{minipage}{14cm}

\begin{center}
\textbf{Abstract}
\end{center}

\vspace*{0.2cm}

Synaptic delays are fundamental determinants of neuronal communication and can profoundly influence the emergence and stability of cortical oscillations. Although their role in shaping network synchronization is well established, how synaptic delays regulate the collective response of neuronal populations to transient perturbations remains poorly understood. Here, we investigate the effects of synaptic delays on the phase and amplitude responses of oscillatory activity in a conductance-based excitatory-inhibitory spiking network operating in the pyramidal-interneuron gamma (PING) regime. By systematically varying the synaptic delay and applying brief external perturbations to the excitatory population, inhibitory population, or the entire network, we computed network phase response curves (nPRCs) and network amplitude response curves (nARCs) to quantify changes in oscillation timing and population coherence. Increasing synaptic delay slowed network oscillations while enhancing population synchrony, demonstrating a trade-off between oscillation frequency and coherence. Excitatory perturbations produced relatively robust phase responses across delays but exhibited a pronounced delay-dependent reduction in amplitude enhancement. In contrast, inhibitory perturbations generated substantially stronger delay-dependent modulation of both phase resetting and amplitude suppression, whereas whole-network stimulation combined features of both excitatory and inhibitory responses. Taken toghether, these findings identify synaptic delay as a key parameter governing the balance between phase resetting and amplitude modulation and provide a computational framework for understanding delay-dependent control of oscillatory brain networks.\\

\textbf{Keywords:} Synaptic delay, oscillation, phase response curve, amplitude response curve, excitatory-inhibitory network model.

\end{minipage}
}
\end{center}


\vspace{0.5cm}

\section{Introduction}

Neuronal oscillations are a fundamental feature of brain activity, arising from the collective and coordinated dynamics of large populations of neurons. Such oscillations have been observed across a wide range of temporal scales and are associated with diverse cognitive and behavioral functions in the brain, including sensory processing, attention, learning and memory, and motor control~\cite{bacsar2001gamma,ward2003synchronous,fries2005mechanism,wang2010neurophysiological}. In fact, oscillatory brain activity reflects the emergence of coherent population rhythms driven by recurrent interactions between excitatory and inhibitory neurons and feedback mechanisms within neuronal circuits~\cite{buzsaki2004neuronal}.

Synchronization in neuronal networks is strongly shaped by inhibition~\cite{van1994inhibition,whittington2000inhibition,tiesinga2009cortical}. Previous studies have shown that inhibitory interactions can promote, stabilize, or disrupt synchrony depending on network architecture, synaptic strengths, time delays and intrinsic neuronal properties~\cite{brunel2000dynamics,marella2010amplification,tetzlaff2012decorrelation,mejias2014differential,hatamian2026modulation}. In particular, excitatory-inhibitory networks are known to generate fast oscillations in gamma band (30-100 Hz) through mechanisms such as pyramidal-interneuron gamma (PING) and interneuron gamma (ING) rhythms~\cite{tiesinga2009cortical}. These oscillations play a centeral role in brain behavior and cognition~\cite{buzsaki2004neuronal}, and arise from the interplay between excitation and delayed inhibition, making the timing of interactions a key control parameter~\cite{madadiasl2025entrainment}. Inhibitory feedback plays a fundamental role in the generation and regulation of oscillatory activity. In particular, fast-spiking interneurons deliver strong and precisely timed synaptic inhibition that can synchronize excitatory populations, control the frequency of network oscillations, and define temporal windows that constrain spike timing and coordination~\cite{bartos2007synaptic,buzsaki2012mechanisms}.

However, excitatory-inhibitory interactions are shaped by synaptic delays, which introduce temporal offsets in neuronal communication and can influence collective network dynamics. Time delays are unavoidable in neuronal systems due to finite axonal conduction speeds, synaptic transmission times, and dendritic processing~\cite{roxin2005role,ermentrout2009delays,madadi19time}. Even relatively small delays can qualitatively alter the influence of inhibition on network dynamics~\cite{maex2003resonant,ledoux2011dynamics,roohi2022role}, leading to changes in oscillation frequency, synchronization strength, and stability~\cite{ernst1995synchronization,crook1997role,brunel2000dynamics,roxin2005role,madadiasl2017dendritic,madadi2018delay}. For example, appropriately timed delays can promote synchronized firing, thereby facilitating the emergence of robust oscillatory states~\cite{dodla2006well,guo2012complex}. Nonetheless, the role of synaptic delays in shaping population responses to perturbations remains less well understood.

Phase response curves (PRCs) provide a powerful framework for characterizing how oscillatory systems respond to transient perturbations~\cite{hansel1995synchrony,ermentrout1996type}. Originally developed for individual oscillators, this concept has been extended to the population level through network phase response curves (nPRCs)~\cite{kawamura2008collective,levnajic2010phase,dumont2017macroscopic}. nPRCs quantify how brief inputs alter the timing of collective rhythms, providing insights into the stability and controllability of synchronized states. However, perturbations can influence not only oscillation timing but also the magnitude and coherence of population activity. To capture these complementary aspects of network responsiveness, network amplitude response curves (nARCs) have been introduced to characterize stimulus-induced changes in oscillation strength and synchrony~\cite{aronson1990amplitude}. In neuronal systems, the joint analysis of phase and amplitude responses offers a more comprehensive view of how external inputs reshape ongoing population dynamics~\cite{smeal2010phase}. Importantly, phase resetting and amplitude modulation need not behave identically, and their relationship may depend on network structure and parameters. In particular, excitatory-inhibitory networks with delayed interactions can introduce additional temporal constraints that differentially shape phase and amplitude responses to transient inputs, potentially giving rise to complex delay-dependent effects on collective synchronization.

In this work, we systematically investigate how synaptic delays shape population phase resetting and amplitude modulation in a sparsely connected excitatory-inhibitory spiking network operating in the PING regime. The model consists of a large-scale random network of conductance-based leaky integrate-and-fire (LIF) neurons. Oscillations emerge through the classical PING mechanism, in which recurrent excitation recruits inhibitory interneurons whose delayed feedback transiently suppresses the excitatory population and resets the oscillatory cycle. Under ongoing Poisson background input, the network exhibits stable gamma-band oscillations characterized by coherent population bursts, irregular single-neuron firing, and a well-defined spectral peak. By systematically varying the interaction synaptic delay in the excitatory-inhibitory feedback loop, we modulate the intrinsic timescale and synchrony of network oscillations while preserving the underlying PING architecture.

Within this framework, we compute nPRCs and nARCs in response to brief, temporally localized perturbations applied selectively to the excitatory population, inhibitory population, or the entire network. We show that synaptic delay regulates not only the frequency and synchrony of population oscillations but also their sensitivity to transient external inputs. Specifically, increasing synaptic delay slows the network rhythm while enhancing population synchrony. Excitatory perturbations produce relatively robust phase resetting but delay-dependent amplitude enhancement, whereas inhibitory perturbations exhibit stronger delay-dependent modulation of both phase and amplitude responses. Perturbations delivered to the entire network combine these complementary effects, revealing how excitatory drive and delayed inhibitory feedback jointly determine collective network dynamics. Finally, by comparing the extrema of the nPRCs and nARCs across stimulation paradigms, we demonstrate that synaptic delay acts as a key control parameter governing the balance between phase resetting, amplitude modulation, and synchronization in cortical-like PING networks.


\section{Methods}
\subsection{Network and neuron model}

\begin{figure}[t!]
\centering
\includegraphics[scale = 0.7]{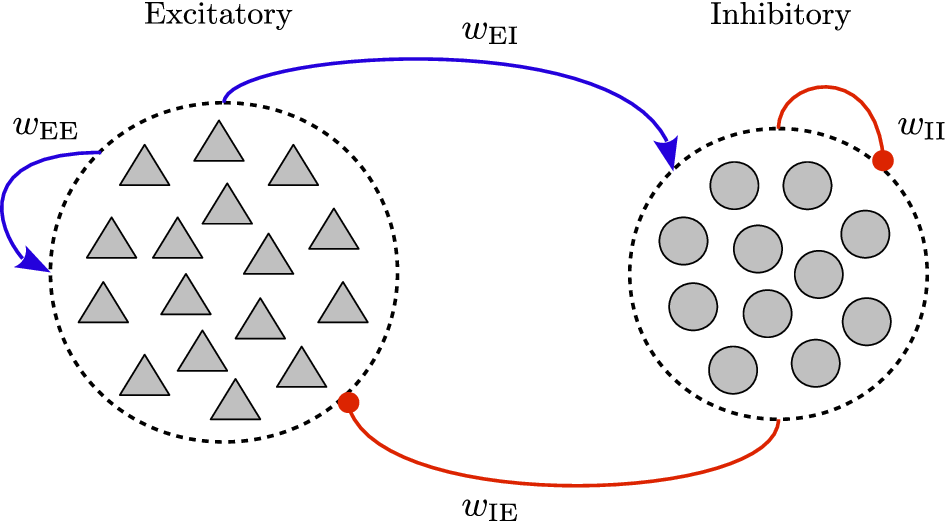}
\caption{{\bf Schematics of the network model.} The model consists of a network of $N_{\mathrm{E}} = 800$ excitatory (triangles) and $N_{\mathrm{I}} = 200$ inhibitory (circles) neurons randomly connected via conductance-based synapses. The excitatory and inhibitory populations in the model interact via synaptic weights $w_{\rm EI}$ (E $\rightarrow$ I) and $w_{\rm IE}$ (I $\rightarrow$ E), characterized by synaptic delays  $d^{\rm EI}_{\rm syn}$ and  $d^{\rm IE}_{\rm syn}$, respectively. Recurrent interactions within each populations are mediated by synaptic weights $w_{\rm EE}$ (E $\rightarrow$ E) in the excitatory population and $w_{\rm II}$ (I $\rightarrow$ I) in the inhibitory population, characterized by synaptic delays  $d^{\rm EE}_{\rm syn}$ and  $d^{\rm II}_{\rm syn}$, respectively.}
\label{fig1}
\end{figure}

As schematically shown in Fig.~\ref{fig1}, we considered a randomly connected excitatory-inhibitory network comprising $N = N_{\mathrm{E}} + N_{\mathrm{I}} = 1000$ neurons, with $N_{\mathrm{E}} = 800$ excitatory and $N_{\mathrm{I}} = 200$ inhibitory neurons in 4:1 proportion. The excitatory and inhibitory populations interacted via synaptic weights $w_{\rm EI}$ (E $\rightarrow$ I) and $w_{\rm IE}$ (I $\rightarrow$ E), characterized by synaptic delays  $d^{\rm EI}_{\rm syn}$ and  $d^{\rm IE}_{\rm syn}$, respectively. Recurrent interactions within each populations were mediated by synaptic weights $w_{\rm EE}$ (E $\rightarrow$ E) in the excitatory population and $w_{\rm II}$ (I $\rightarrow$ I) in the inhibitory population, characterized by synaptic delays  $d^{\rm EE}_{\rm syn}$ and  $d^{\rm II}_{\rm syn}$, respectively. Random and sparse connections between neurons were established with a connection probability of $p = 0.1$.

The neurons were simulated by the LIF neuron model with conductance-based synapses~\cite{burkitt2006review1,burkitt2006review2}, within which the dynamics of the membrane potential ($V_{\rm m}$) of a neuron in the network are described by the following differential equation:
\begin{equation}\label{eq:1}
C_{\rm m} \frac{dV_{\rm m}}{dt} = - g_{\rm L} (V_{\rm m} - E_{\rm L}) + I_{\rm syn}(t),
\end{equation}
where $C_{\rm m}$ is the membrane capacitance, $g_{\rm L}$ is the leak conductance, and $E_{\rm L}$ is the leak reversal potential. The model includes a hard threshold for spike emission ($V_{\rm th}$), a fixed refractory period ($t_{\rm ref}$), and no adaptation mechanisms. A spike is emitted at time step $t^* = t_{k + 1}$ if $V_{\rm m}(t_k) < V_{\rm th}$ and $V_{\rm m}(t_{k + 1}) > V_{\rm th}$. Following spike emission, the membrane potential resets to $V_{\rm{reset}}$ for $t^* < t <  t^* + t_{\rm ref}$, i.e., the membrane potential is clamped to $V_{\rm{reset}}$ during the refractory period. The numerical values of model parameters are listed in Table~\ref{table1}.


\subsection{Synapse model}

In the model, neurons were coupled in a random topology via conductance-based synapses, with synaptic interactions characterized by the synaptic current $I_{\rm syn}(t)$ in Eq.~(\ref{eq:1}). The total synaptic current received by a neuron comprised excitatory and inhibitory components, i.e., $I_{\rm syn}(t) = I^{\rm ex}_{\rm syn}(t) + I^{\rm in}_{\rm syn}(t)$, where each component is given by:
\begin{equation}\label{eq:2}
I^{\rm X}_{\rm syn}(t) = (E^{\rm X}_{\mathrm{syn}} - V_{\rm m}(t)) \sum_{i} \sum_{f} g_{i \rm X}(t - t^{(f)}_i - d_{\rm syn}),
\end{equation}
where $i$ indicates excitatory (X = E) or inhibitory (X = I) presynaptic neurons characterized by the membrane potential $V_{\rm m}(t)$, $E^{\rm X}_{\mathrm{syn}}$ is the corresponding synaptic reversal potential, $g(t)$ is the synaptic conductance, $t^{(f)}$ represents the spike times of the presynaptic neuron, and $d_{\rm syn}$ is the synaptic delay. Each incoming spike produces a postsynaptic conductance change described by an exponential function. The resulting synaptic conductances are given by:
\begin{equation}\label{eq:3}
g_{i \rm X}(t) = w_i \exp \left(  -\dfrac{t}{\tau^{\rm X}_{\rm{syn}}} \right)  \Theta(t),
\end{equation}
where $w$ is the synaptic weight, $\tau^{\rm X}_{\rm{syn}}$ is the synaptic time constant, and $\Theta(t)$ denotes the Heaviside step function, defined as $\Theta(t) = 1$ for $t \geq 0$, and $\Theta(t) = 0$ otherwise. The synaptic conductances are normalized to unit maximum~\cite{meffin2004analytical}, i.e., $g (t = 0) = w$.


\subsection{External background input}

Ongoing background activity was modeled using independent Poisson spike trains. Each excitatory and inhibitory neuron received background inputs with constant firing rates of 1500 spikes/s and 1600 spikes/s, respectively, representing uncorrelated synaptic bombardment from external sources. Incoming background spikes generated transient changes in synaptic conductance with an exponential decay profile, consistent with conductance-based synaptic dynamics. These background inputs provided a fluctuating baseline drive to the neurons, mimicking variable discharge of cortical neurons observed experimentally~\cite{shadlen1998variable}.


\subsection{Phase and amplitude responses to external stimuli}

The PRC of a periodically spiking neuron characterizes how brief perturbations affect the timing of subsequent spikes~\cite{hansel1995synchrony,ermentrout1996type}. In practice, a small current pulse is injected into the neuron, and the resulting shifts in spike timing or phase are measured. To assess the response of the excitatory-inhibitory network to external perturbations in our model, we employed the nPRC framework~\cite{kawamura2008collective,levnajic2010phase,dumont2017macroscopic}, which generalizes the single-neuron PRC to the population level. The nPRC quantifies the collective phase shift of network oscillations in response to external stimuli, providing a macroscopic measure of the network's sensitivity to perturbations.

To quantify the nPRC, we first adjusted the model parameters so that the unperturbed network exhibited synchronized activity. The collective oscillation cycle ($T_0$) was estimated by measuring the interval between two successive peaks of the network oscillation. The network was then perturbed by delivering an excitatory (positive) Gaussian pulse packet - containing 10 temporally localized spikes with times drawn from a normal distribution (standard deviation 0.1 ms) - to neurons at different phases ($\theta$) of the oscillation cycle. The resulting phase shifts of the network oscillation were subsequently measured:
\begin{equation}\label{eq:4}
{\rm nPRC}(\theta) = 1 - \dfrac{T^{\prime}}{T_0},
\end{equation}
where $\theta = \frac{2 \pi}{T_0} (t  - t_n)$ is the phase of network oscillation, $T^{\prime} = t_{n + 1} - t_{n}$ is the oscillation cycle of the perturbed network, $t_{n}$ is the time of the last synchronous population peak before perturbation and $t_{n + 1}$ is the time of the first synchronous population peak after perturbation that could be advanced or delayed.

To quantify the network's amplitude response, we computed the nARC~\cite{aronson1990amplitude}, providing a measure of how temporally localized perturbations modulate the magnitude of network oscillations at the population level. Specifically, for each perturbation, we measured the change in amplitude caused by the perturbation:
\begin{equation}\label{eq:5}
{\rm nARC}(\theta) = \dfrac{A^{\prime}}{A_0} - 1,
\end{equation}
where $A_0 = A(t_n)$ is the unperturbed peak amplitude and $A^{\prime} = A(t_{n + 1})$ is the amplitude of the first synchronous population peak after perturbation.

\begin{table}[t!]
\centering
\caption{{\bf Parameters of the network and neuron model used in our simulations.}}
\vspace*{-0.2cm}
\begin{tabular}{|l|c|c|c|}
\hline
{\bf Parameter} & {\bf Symbol} & {\bf Value} & {\bf Unit}\\ \hline
Membrane capacitance & $C_{\rm m}$ & 250 & $ \mathrm{pF}$\\ \hline
Spiking threshold & $V_{\mathrm{th}}$ & -55 & $\mathrm{mV}$\\ \hline
Resting membrane potential & $V_{\mathrm{reset}}$ & -70 & $\mathrm{mV}$\\ \hline
Leak reversal potential & $E_{\rm L}$ & -70 & $\mathrm{mV}$\\ \hline
Leak conductance & $g_{\rm L}$ & 16.7 & $\mathrm{nS}$\\ \hline
Duration of refractory period & $t_{\rm ref}$ & 2 & $\mathrm{ms}$\\ \hline
Excitatory reversal potential & $E^{\mathrm{ex}}_{\rm syn}$ & 0 & $\mathrm{mV}$\\ \hline
Inhibitory reversal potential & $E^{\mathrm{in}}_{\rm syn}$ & -75 & $\mathrm{mV}$\\ \hline
Decay time constant of excitatory synapses & $\tau^{\rm ex}_{\rm{syn}}$ & 2 & $\mathrm{ms}$\\ \hline
Decay time constant of inhibitory synapses & $\tau^{\rm in}_{\rm{syn}}$ & 3 & $\mathrm{ms}$\\ \hline
${\rm E} \rightarrow {\rm E}$ synaptic delay & $d^{\rm{EE}}_{\rm syn}$ & 0.1 & $\mathrm{ms}$\\ \hline
${\rm I} \rightarrow {\rm I}$ synaptic delay & $d^{\rm{II}}_{\rm syn}$ & 0.1 & $\mathrm{ms}$\\ \hline
${\rm E} \rightarrow {\rm E}$ synaptic weight & $w_{\rm{EE}}$ & 1 & $\mathrm{nS}$\\ \hline
${\rm E} \rightarrow {\rm I}$ synaptic weight & $w_{\rm{EI}}$ & 10 & $\mathrm{nS}$\\ \hline
${\rm I} \rightarrow {\rm E}$ synaptic weight & $w_{\rm{IE}}$ & -12 & $\mathrm{nS}$\\ \hline
${\rm I} \rightarrow {\rm I}$ synaptic weight & $w_{\rm{II}}$ & -13 & $\mathrm{nS}$\\ \hline
Number of excitatory neurons & $N_{\mathrm{E}}$ & 800 & -- \\ \hline
Number of inhibitory neurons & $N_{\mathrm{I}}$ & 200 & -- \\ \hline
Total number of neurons & $N_{\mathrm{I}}$ & 1000 & -- \\ \hline
Connection probability & $p$ & 0.1 & -- \\ \hline
\end{tabular}
\label{table1}
\end{table}


\subsection{Analysis of the network dynamics}
\subsubsection{Population activity}

The activity of the network was calculated by counting the number of spikes in a time interval which gives the number of active neurons at that interval as an indicator of synchronized dynamics, excluding the first 200 ms of transients:
\begin{equation}\label{eq:6}
A_{\rm X}(t) = \dfrac{1}{N_{\rm X}} \sum^N_{i=1} \sum_f \delta(t - t_i^{(f)}),
\end{equation}
where $N_{\rm X}$ is the total number of neurons in the excitatory ($N_{\mathrm{E}} = 800$) or inhibitory ($N_{\mathrm{I}} = 200$) subpopulation, and $t^{(f)}$ represents the firing time of individual neurons. $A(t)$ is reported as the percentage of active neurons.


\subsubsection{Synchrony index}

The population Fano factor (pFF) was used to estimate the synchrony of population activity in the network, excluding the first 200 ms of transients~\cite{kumar2008conditions}:
\begin{equation}\label{eq:7}
{\mathrm{pFF}} = \dfrac{\sigma^2 [A(t)]}{\mu [A(t)]},
\end{equation}
where $A(t)$ represents the population activity defined in Eq.~(\ref{eq:6}), and $\sigma^2$ and $\mu$ are the variance and mean of the population activity, respectively. The pFF evaluates the normalized amplitude of the variation in the network activity which increases when neurons fire in synchrony~\cite{kumar2008conditions,madadi2023decoupling}. Smaller values of the pFF correspond to desynchronized states, whereas greater values of the pFF imply synchrony in the network.


\subsubsection{Spike count irregularity}

The coefficient of variation (CV) of the inter-spike intervals (ISIs) was calculated as a measure of the irregularity of the spiking activity of neuron $i$, as follows:
\begin{equation}\label{eq:8}
{\mathrm{CV}_i} = \dfrac{\sigma_i}{\mu_i},
\end{equation}
where $\sigma_i$ is the standard deviation and $\mu_i$ is the mean of the ISIs calculated from the spike time distribution of neuron.


\subsubsection{Network frequency and power spectrum}

The network oscillation frequency was defined as the peak frequency of the power spectral density (PSD) of the population activity. To estimate the PSD, we applied Welch's method~\cite{welch1967use}, dividing the population activity time series into overlapping segments with 50\% overlap and using a sampling frequency of 1000 Hz. The resulting PSD allowed us to identify the dominant oscillatory frequency of the network under different simulation conditions.


\subsection{Model integration}

Computer simulations were performed using the NEural Simulation Tool (NEST)~\cite{gewaltig2007nest}. The simulation code and the associated data used to generate the figures are publicly available at \href{https://github.com/parsa-shr/Synaptic-delay-phase-amplitude-response}{https://github.com/parsa-shr/Synaptic-delay-phase-amplitude-response}. The subthreshold dynamics in the model were integrated using the \texttt{iaf\_cond\_exp} model implemented in NEST to simulate a random and sparsely connected network of spiking LIF neurons with conductance-based synapses, utilizing an integration time step of 0.01 ms. For each numerical experiment, the model was run for $1.0 \times 10^5$ time steps, resulting in a simulation duration of 1000 ms. The simulation results are displayed within a time window appropriate for each specific figure. Data analysis was performed using tools implemented in NEST.


\section{Results}
\subsection{PING oscillations and baseline network dynamics}

\begin{figure}[t!]
\centering
\includegraphics[scale = 0.72]{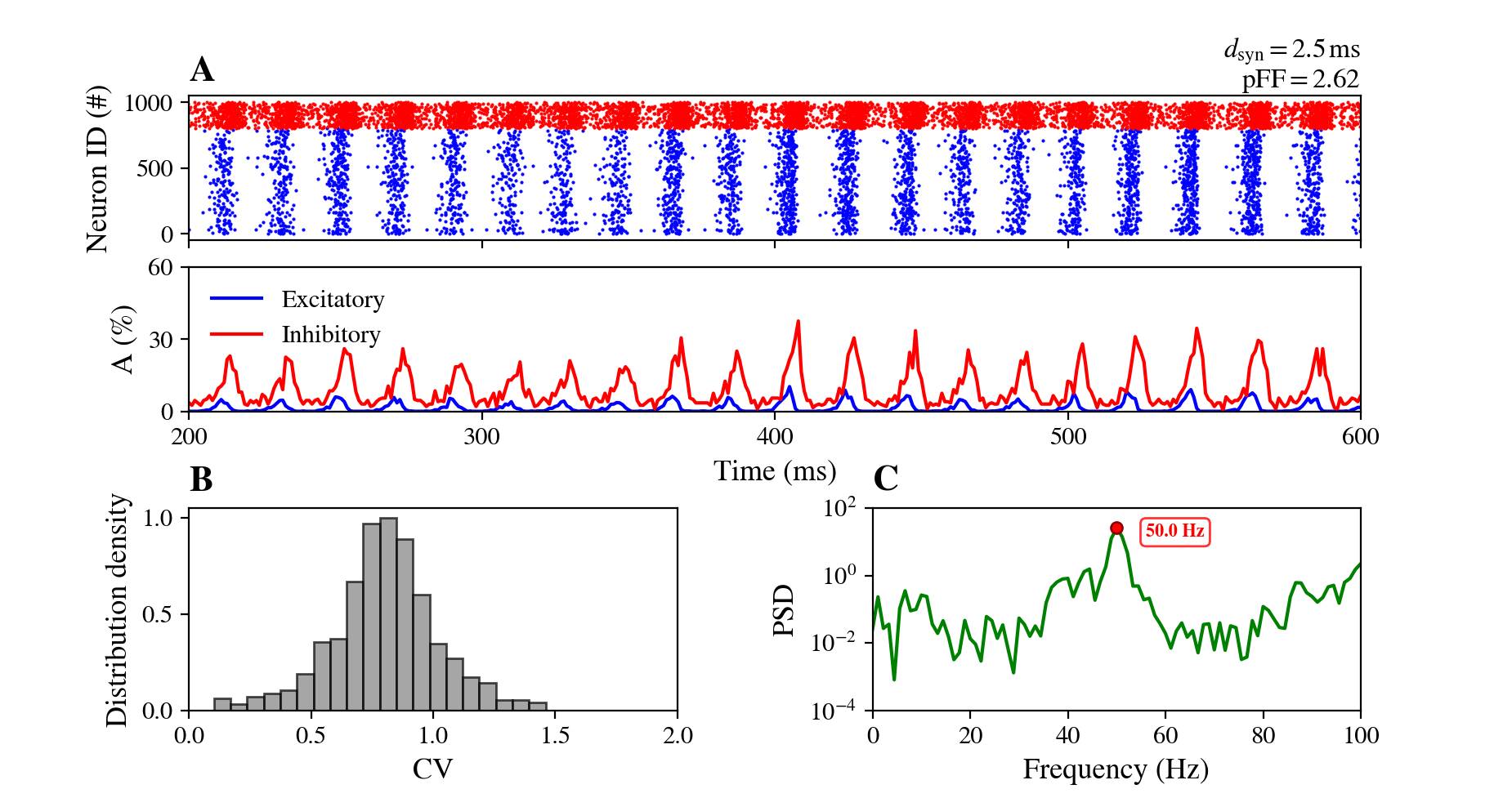}
\caption{{\bf Dynamics of the excitatory-inhibitory network model.} (\textbf{A}) Raster plot (top) and corresponding population activity (bottom) of the network of excitatory (blue) and inhibitory (red) neurons, exhibiting self-sustained oscillations characteristic of the PING mechanism. Model parameters are given in Table~\ref{table1}, with $d_{\rm syn} = 2.5 \, \mathrm{ms}$. The level of synchrony (assessed by calculating pFF) is indicated above the panel (pFF = 2.62). (\textbf{B}) Spike count irregularity of individual neurons (assessed by calculating CV). (\textbf{C}) PSD of the population activity, showing a prominent oscillatory peak at approximately 50 Hz.}
\label{fig2}
\end{figure}

To characterize the baseline dynamics, we simulated the excitatory-inhibitory network described in Methods. Synaptic delays between the excitatory and inhibitory populations were assumed to be identical ($d^{\rm EI}_{\rm syn} = d^{\rm IE}_{\rm syn} = d_{\rm syn}$) and were systematically varied to examine the resulting network dynamics. For a synaptic delay of $d_{\rm syn} = 2.5 \, \rm ms$, for instance, the excitatory-inhibitory network exhibits robust, self-sustained oscillations characteristic of the PING mechanism. As shown in Fig.~\ref{fig2}A (top; raster plot), the excitatory (blue) and inhibitory (red) populations organize into rhythmic spike clusters, indicating strong population-level coordination, as reflected by a pFF of 2.62. The rhythmic alternation between excitatory and inhibitory spiking is visible in the bursts of excitatory activity followed by synchronized inhibitory firing, consistent with the classical PING cycle in which excitation recruits inhibition, and inhibition in turn suppresses excitation~\cite{tiesinga2009cortical}. The corresponding population activity traces (Fig.~\ref{fig2}A, bottom) further confirm the presence of coherent oscillations. The excitatory activity rises sharply at the onset of each cycle and is followed, with a short delay, by a larger inhibitory peak.

The spike time variability of individual neurons, quantified by the CV distribution (Fig.~\ref{fig2}B) indicates irregular spiking at the single-neuron level despite strong macroscopic synchronization. This coexistence of irregular microscopic dynamics and coherent macroscopic oscillations is a well-known property of balanced excitatory-inhibitory networks~\cite{shadlen1998variable}. Such fluctuation-driven yet rhythmically organized activity is typical of PING networks operating in a sparsely connected regime. Spectral analysis of the population activity (Fig.~\ref{fig2}C) reveals a pronounced peak at approximately 50 Hz, placing the oscillation firmly within the gamma frequency band (30-100 Hz), which is a hallmark of coherent PING oscillations. In this regime, the oscillation frequency reflects the combined effects of excitatory recruitment, inhibitory synaptic time constants, and the synaptic delay.

Overall, the simulated network operates in a classical PING regime characterized by (i) sequential excitatory-inhibitory activation, (ii) stable gamma-band oscillations, (iii) coherent population rhythms with irregular single-neuron firing, and (iv) frequency determined by the interaction between synaptic kinetics and delay. This dynamical state provides a well-defined baseline for investigating how transient perturbations and delay variations modulate collective phase and amplitude responses in the subsequent analyses.


\subsection{Delay-dependent changes in oscillatory dynamics}

\begin{figure}[t!]
\centering
\includegraphics[scale = 0.72]{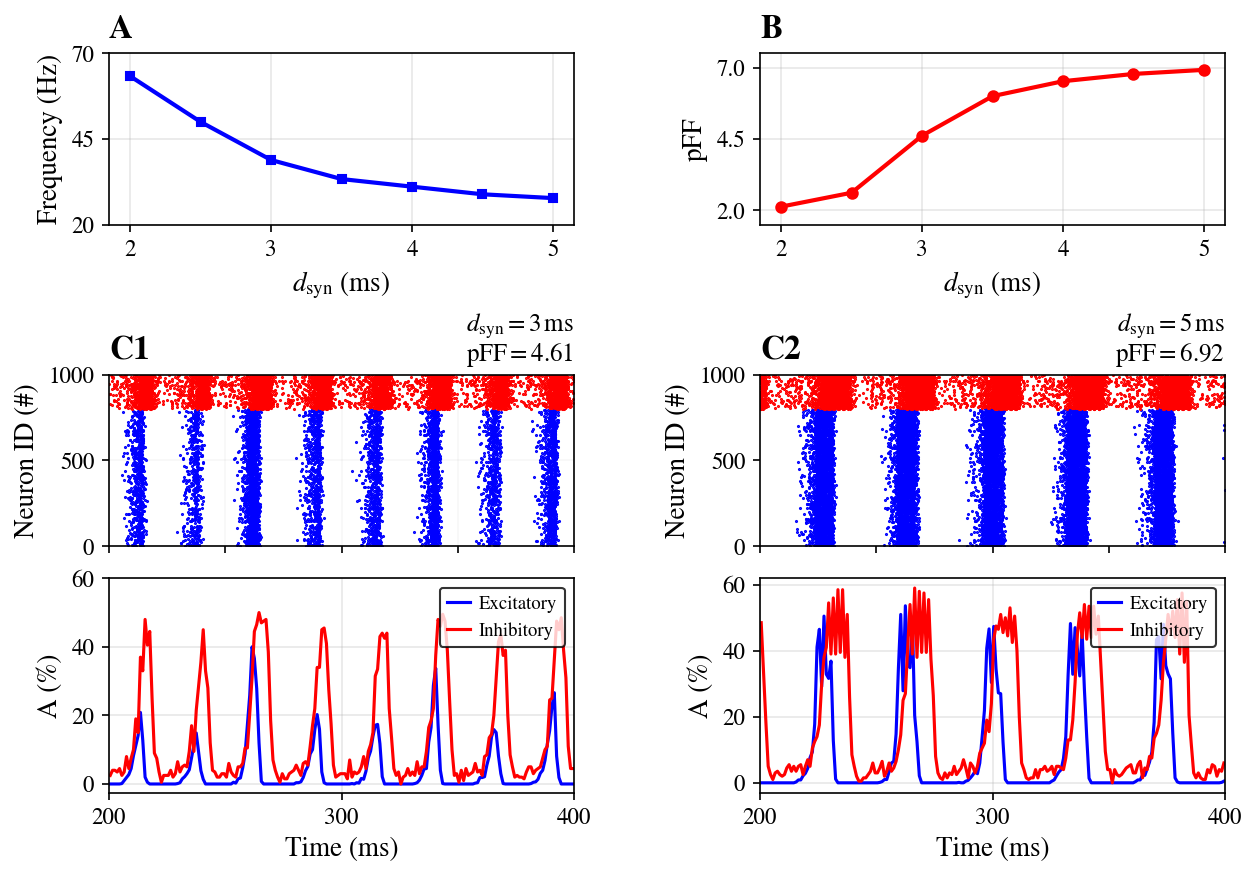}
\caption{{\bf Delay-dependent changes in oscillation frequency and synchrony.} (\textbf{A}) The peak frequency of the PSD as a function of synaptic delays. (\textbf{B}) The level of synchrony (assessed by calculating pFF) for different synaptic delays. (\textbf{C1, C2}) Representative raster plot (top) and corresponding population activity (bottom) of the network for $d_{\rm syn} = 3.0 \, \mathrm{ms}$ (C1) and $d_{\rm syn} = 5.0 \, \mathrm{ms}$ (C2). pFF is indicated above each panel with pFF = 4.61 for panel C1 and pFF = 6.92 for panel C2.}
\label{fig3}
\end{figure}

Systematic variation of the synaptic delay reveals a pronounced impact on both the oscillation frequency (assessed by calculating PSD peak) and the degree of population synchrony (assessed by calculating pFF) in the excitatory-inhibitory network. As shown in Fig.~\ref{fig3}A, increasing the delay from 2.0 ms to 5.0 ms leads to a monotonic decrease in the dominant oscillation frequency. The network transitions from fast gamma oscillations near $\sim 60$ Hz at shorter delays to slower rhythms approaching $\sim 30$ Hz at longer delays. This reduction in frequency reflects the fundamental role of delayed inhibition in setting the oscillatory timescale: longer delays prolong the excitatory-inhibitory feedback loop, effectively extending the duration of each oscillatory cycle. Since PING rhythms are governed by the sequential activation of excitatory and inhibitory populations, increasing the delay directly increases the period of the E $\rightarrow$ I $\rightarrow$ E loop, thereby lowering the frequency.

While the oscillation frequency decreases with delay, population synchrony exhibits the opposite trend (Fig.~\ref{fig3}B). The pFF increases substantially as synaptic delay grows, indicating stronger fluctuations in population activity and thus enhanced synchronization. Larger pFF values (e.g., pFF = 6.92 for $d_{\rm syn} = 5.0 \, \mathrm{ms}$) correspond to more coherent spike clustering across neurons, suggesting that longer delays promote tighter temporal alignment of firing events. This behavior indicates that delayed inhibition does not merely slow down oscillations, but also strengthens the collective coordination of neuronal activity.

The raster plots and population activity traces further illustrate these delay-dependent changes. For $d_{\rm syn} = 3.0 \, \mathrm{ms}$ (Fig.~\ref{fig3}C1), the network displays clear rhythmic spike packets, but the excitatory clusters remain relatively broad and less sharply defined. The corresponding population activity peaks are moderate in amplitude and somewhat narrower in temporal separation (pFF = 4.61), consistent with the higher oscillation frequency observed at shorter delays (i.e., 38.9 Hz for $d_{\rm syn} = 3.0 \, \mathrm{ms}$). In contrast, for $d_{\rm syn} = 5.0 \, \mathrm{ms}$ (Fig.~\ref{fig3}C2), spike clusters become more compact and temporally aligned, reflecting stronger synchrony. The population activity peaks are larger in amplitude and more widely spaced, consistent with both increased synchrony (pFF = 6.92) and reduced oscillation frequency (27.8 Hz).

Mechanistically, the increase in synchrony with longer delays can be understood in terms of the inhibitory reset dynamics. A longer delay allows excitatory neurons to accumulate activity before inhibition arrives, leading to a more collective and abrupt suppression of firing. This synchronized inhibitory feedback effectively resets a larger fraction of the excitatory population simultaneously, producing sharper and more coherent population bursts in the subsequent cycle. As a result, oscillations become slower but more synchronized. Taken together, these results demonstrate a clear trade-off controlled by synaptic delay: increasing $d_{\rm syn}$ slows the network rhythm while simultaneously enhancing population synchrony. This dual effect underscores the critical role of synaptic delays in shaping the macroscopic dynamics of PING networks, highlighting delay as a key control parameter for tuning both oscillation frequency and coherence in excitatory-inhibitory circuits.


\subsection{Network response to external perturbations}

\begin{figure}[t!]
\centering
\includegraphics[scale = 0.42]{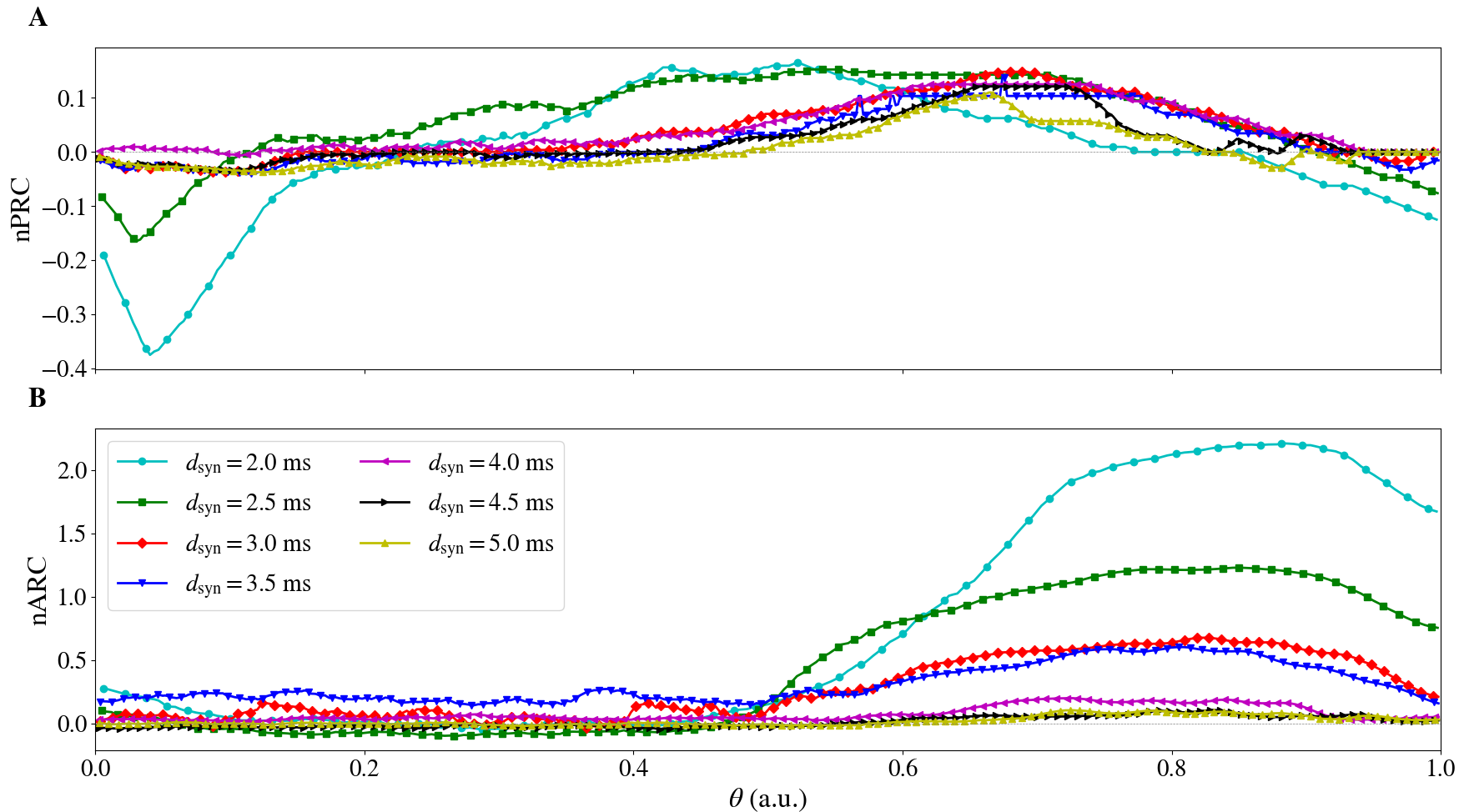}
\caption{{\bf Phase and amplitude responses to external stimulation of the excitatory population.} The nPRC (\textbf{A}) and nARC (\textbf{B}) of the excitatory-inhibitory network obtained by perturbing the excitatory population with different synaptic delays in synchronized state (see Methods). The gray dashed lines indicate zero response.}
\label{fig4}
\end{figure}

To investigate the phase and amplitude responses of the excitatory-inhibitory network to external perturbations for different synaptic delays, the network in synchronized state was perturbed with a short positive spike train consisting of temporally localized Gaussian pulse packets, and the resultant changes in network oscillation cycle and amplitude were calculated (see Methods). The findings are concluded in Figs.~\ref{fig4}, \ref{fig6}, and \ref{fig7} where only the excitatory population (Fig.~\ref{fig4}), only the inhibitory population (Fig.~\ref{fig6}), and the whole excitatory-inhibitory network (Fig.~\ref{fig7}) with different synaptic delays were perturbed.

Fig.~\ref{fig4} shows the resulting nPRC (A) and nARC (B) as functions of the oscillation phase ($\theta$, normalized to unity) for different synaptic delays ($d_{\rm syn} = 2.0$-$5.0 \, \rm ms$), when only the excitatory population was perturbed by external stimulation. In general, excitatory perturbations induced phase advances and amplitude enhancements during the latter half of the oscillation cycle ($\theta > 0.5$), reflecting the increased excitatory drive delivered close to the phase at which the network is approaching the next population burst. For short synaptic delays, the nPRC exhibited a biphasic structure, with a negative component preceding a positive lobe, characteristic of a type-II-like collective phase response. As the synaptic delay increased, the negative component progressively diminished, yielding a predominantly positive response that resembled a type-I collective phase response.

In contrast, the magnitude of amplitude modulation exhibited a pronounced dependence on synaptic delay (notably for $\theta > 0.5$). Shorter delays produced substantially larger nARC responses, with the strongest increase observed for $d_{\rm syn} = 2.0 \, \rm ms$, whereas longer delays progressively reduced the amplitude modulation. This behavior suggests that the impact of an excitatory perturbation depends on the timing relationship between the external input and the intrinsic excitatory-inhibitory feedback loop. When inhibition arrives relatively early, an additional excitatory drive can effectively recruit a larger fraction of the excitatory population before inhibitory suppression, resulting in a stronger transient enhancement of population activity. As the synaptic delay increases, the network becomes more strongly synchronized and dominated by delayed inhibitory feedback, reducing the relative effect of an additional excitatory perturbation on oscillation amplitude.

\begin{figure}[t!]
\centering
\includegraphics[scale = 0.6]{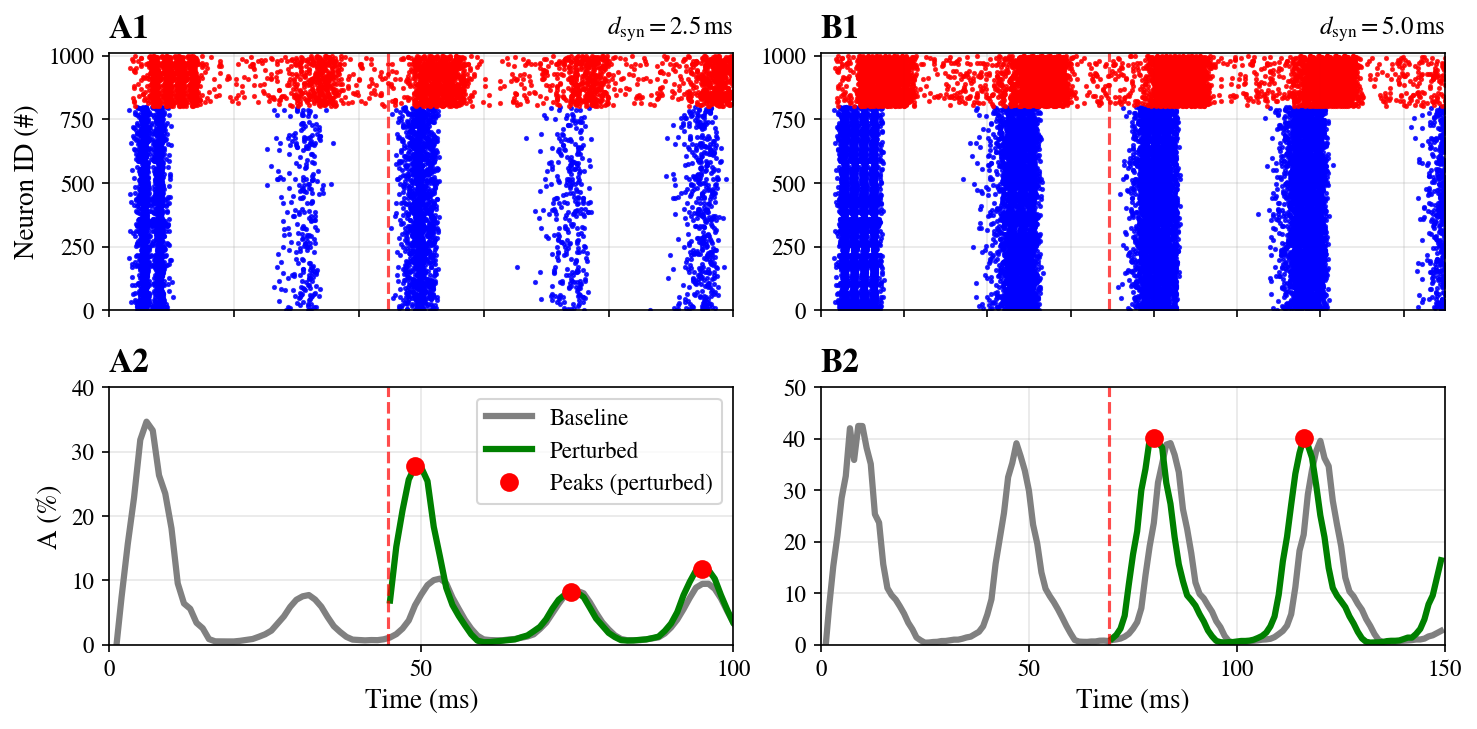}
\caption{{\bf Representative response of the excitatory-inhibitory network to excitatory population perturbation.} Raster plots (\textbf{A1}, \textbf{B1}) and population activity (\textbf{A2}, \textbf{B2}) responses following a brief excitatory pulse applied to the excitatory population at a selected phase of the oscillation cycle for a shorter synaptic delay ($d_{\rm syn}=2.5~\rm ms$), and a longer synaptic delay ($d_{\rm syn}=5.0~\rm ms$). Raster plots show excitatory (blue) and inhibitory (red) neuronal activity before (baseline) and after the perturbation (red dashed line indicates the pulse timing). Population activity traces before (gray) and after (green) perturbation show the phase shift and amplitude modulation of subsequent oscillatory cycles. The perturbed oscillation peaks are indicated by red markers. The results demonstrate that excitatory stimulation can advance population rhythms and enhance oscillation amplitude through transient recruitment of excitatory neurons within the PING feedback loop.}
\label{fig5}
\end{figure}

\begin{figure}[t!]
\centering
\includegraphics[scale = 0.35]{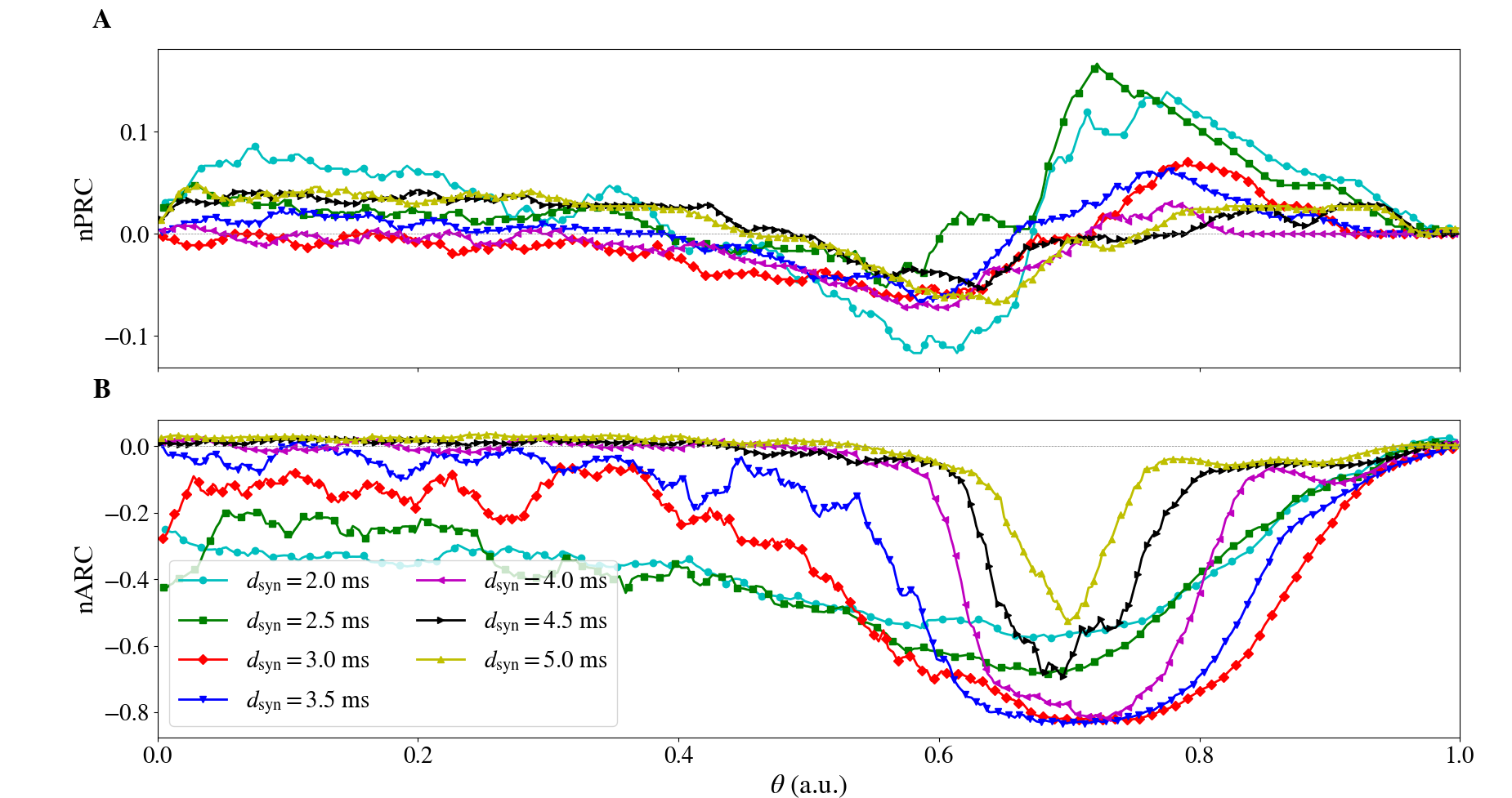}
\caption{{\bf Phase and amplitude responses to external stimulation of the inhibitory population.} The nPRC (\textbf{A}) and nARC (\textbf{B}) of the excitatory-inhibitory network obtained by perturbing the inhibitory population with different synaptic delays in synchronized state. The gray dashed lines indicate zero response.}
\label{fig6}
\end{figure}

\begin{figure}[t!]
\centering
\includegraphics[scale = 0.42]{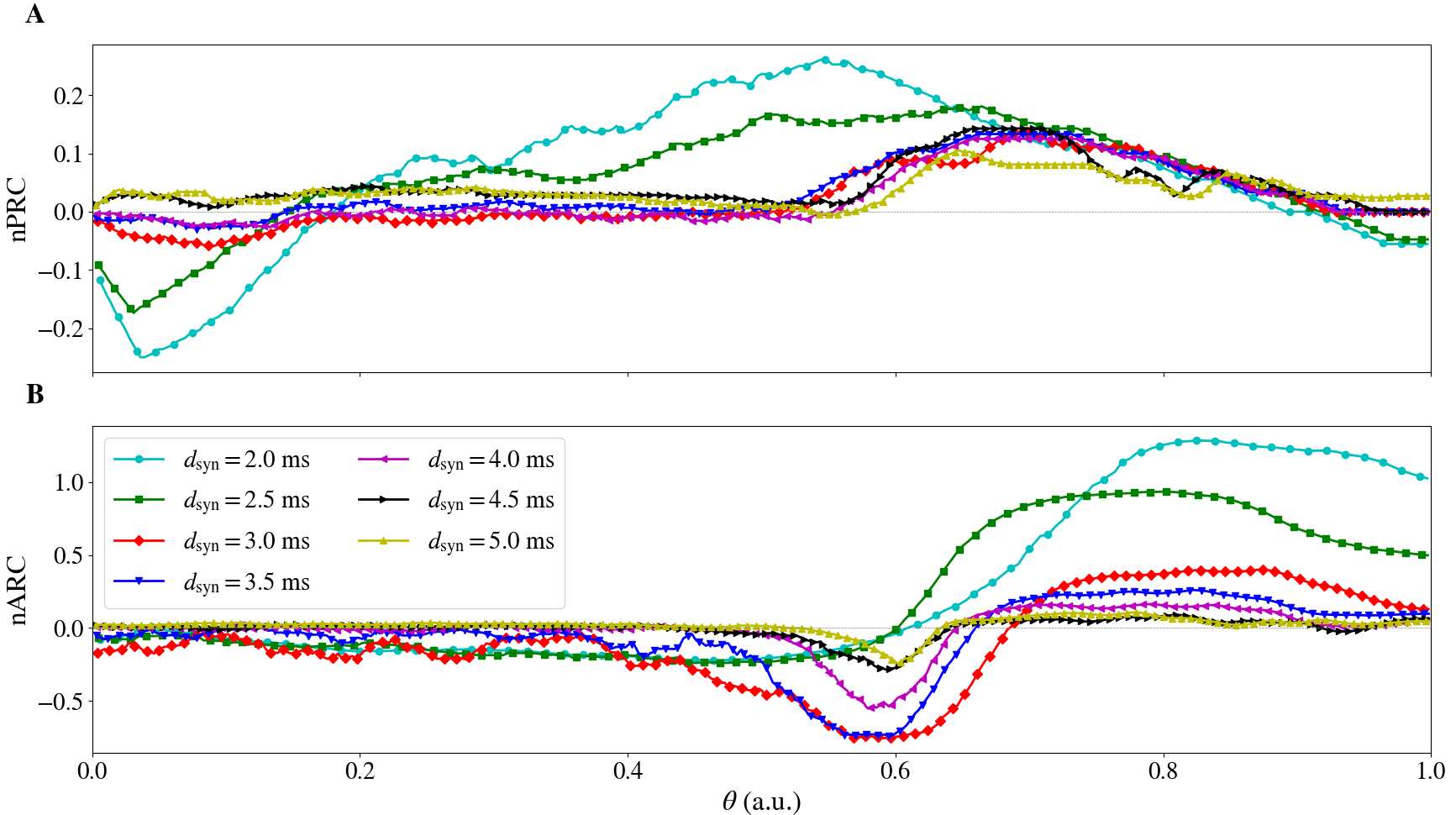}
\caption{{\bf Phase and amplitude responses to external stimulation of the whole excitatory-inhibitory network.} The nPRC (\textbf{A}) and nARC (\textbf{B}) of the excitatory-inhibitory network obtained by perturbing the whole network with different synaptic delays in synchronized state. The gray dashed line indicates zero response.}
\label{fig7}
\end{figure}

A representative example of the effect of an excitatory perturbation on network dynamics is shown in Fig.~\ref{fig5}. For both shorter ($d_{\rm syn} = 2.5 \, \mathrm{ms}$; left column) and longer ($d_{\rm syn} = 5.0 \, \mathrm{ms}$; right column) synaptic delays, the perturbation induces a transient increase in excitatory activity followed by a modified excitatory-inhibitory sequence. The raster plots (A1 and B1) and population activity traces (A2 and B2) demonstrate that a brief perturbation delivered at a specific phase of the oscillation cycle (red dashed line) advances the timing of subsequent population bursts and enhances the amplitude of the following oscillatory cycle (Fig.~\ref{fig5}A2 and B2; cf. gray and green curves). These results illustrate how transient excitation interacts with the intrinsic delayed inhibitory feedback mechanism to reshape both the timing and strength of collective network oscillations.

In contrast, when perturbations were applied to the inhibitory population, the network responses exhibited stronger dependence on synaptic delay, as shown in Fig.~\ref{fig6}. The nPRC (A) displayed a delay-dependent reshaping of the phase response, particularly during the second half of the oscillation cycle ($\theta > 0.5$). For small phases, inhibitory stimulation produced relatively weak phase shifts with a dominant positive response during the late phase of the cycle, resembling a type-I collective response. As the synaptic delay increased, the nPRC developed a more pronounced biphasic structure, with a negative component preceding a positive lobe, indicating a transition toward type-II-like collective phase resetting. This behavior suggests that delayed inhibitory feedback modifies the phase sensitivity of the population by altering the temporal relationship between inhibitory input and the intrinsic excitatory-inhibitory oscillatory cycle. The amplitude responses to inhibitory perturbations showed a distinct behavior from the phase responses (Fig.~\ref{fig6}B). For shorter synaptic delays, inhibitory stimulation primarily reduced the amplitude of population oscillations, as reflected by predominantly negative nARC values. However, the magnitude and timing of amplitude suppression were strongly delay-dependent. Intermediate delays produced the largest reduction in oscillation amplitude, whereas longer delays resulted in weaker and more temporally localized suppression. These results indicate that inhibitory perturbations can transiently disrupt population coherence, with the extent of desynchronization determined by the timing of inhibitory feedback relative to the ongoing oscillatory cycle. Inhibitory perturbations therefore exhibit substantially greater delay sensitivity than excitatory perturbations.

When the entire excitatory-inhibitory network was perturbed simultaneously (Fig.~\ref{fig7}), the collective response reflected the combined effects of excitatory drive and inhibitory modulation. The nPRC (A) exhibited a predominantly type-I-like response across different delays, with a dominant positive phase shift during the latter half of the oscillation cycle. However, the magnitude and shape of the phase response remained delay-dependent. In contrast, the nARC (B) showed a more complex delay-dependent modulation, with shorter delays producing a strong positive amplitude response (notably for $\theta > 0.5$), whereas longer delays resulted in weaker enhancement or nearly unchanged oscillation amplitude. Together, these findings demonstrate that synaptic delays critically regulate how excitatory and inhibitory perturbations reshape collective oscillations, influencing both phase resetting and amplitude modulation in PING networks. Whole-network stimulation integrates these complementary mechanisms.

\subsection{Summary of delay-dependent responses}

\begin{figure}[t!]
\centering
\includegraphics[scale = 0.4]{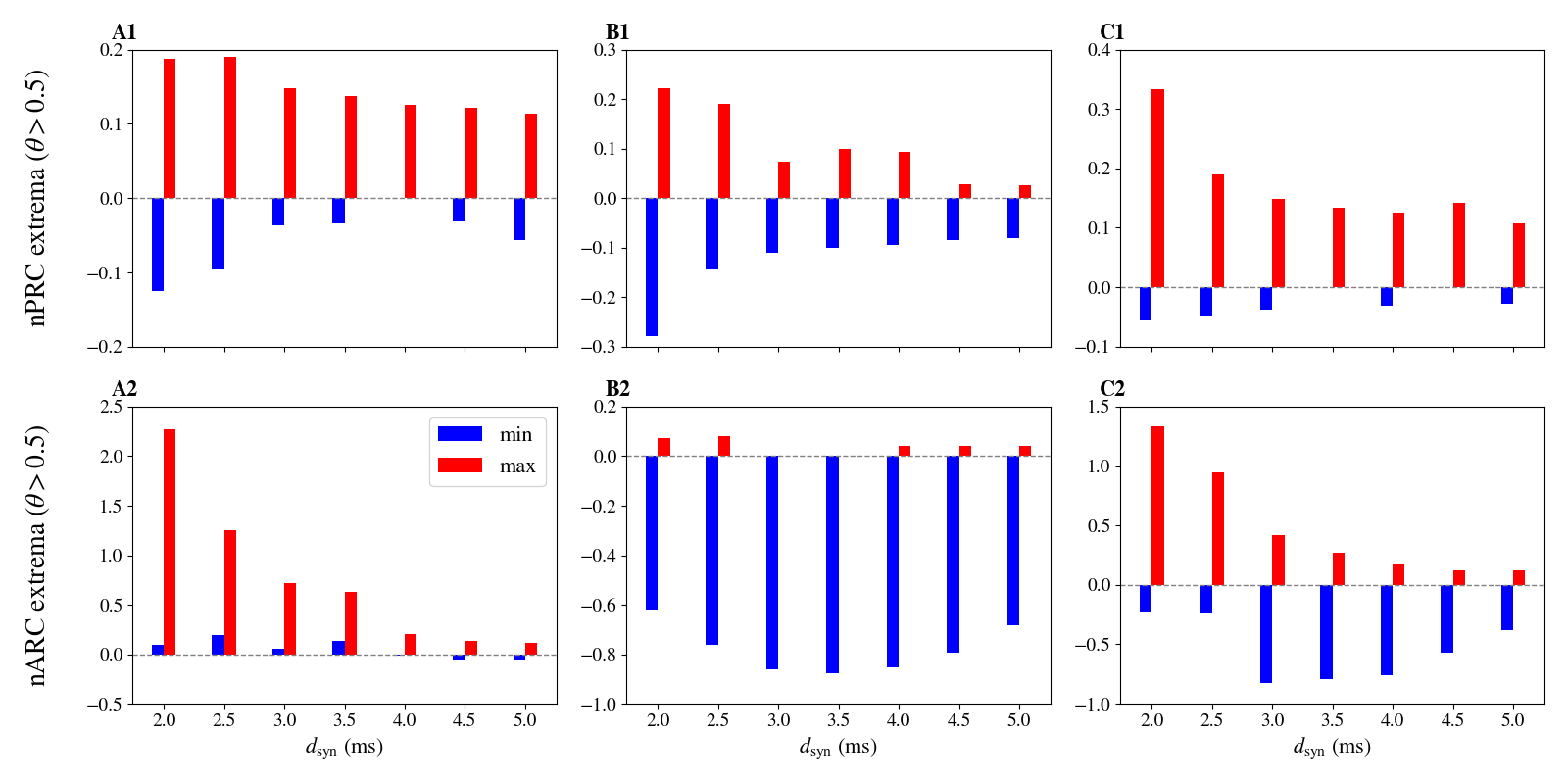}
\caption{{\bf Summary of delay-dependent changes in population phase and amplitude modulations.} Extrema of the nPRC (top row; \textbf{A1-C1}) and nARC (bottom row; \textbf{A2-C2}) for $\theta > 0.5$ as a function of synaptic delay for excitatory (left column; A1 and A2), inhibitory (middle column; B1 and B2), and whole-network (right column; C1 and C2) stimulation. The maximum and minimum response values quantify how synaptic delays regulate the strength and polarity of perturbation-induced changes in population oscillations. The gray dashed lines indicate zero response. This summary highlights synaptic delay as a key parameter controlling the phase sensitivity and amplitude stability of collective oscillations in excitatory-inhibitory networks.}
\label{fig8}
\end{figure}

To facilitate direct comparison between stimulation paradigms, Fig.~\ref{fig8} summarizes the extrema (maximum and minimum values) of the nPRCs and nARCs obtained for excitatory, inhibitory, and whole-network perturbations for $\theta > 0.5$ where the main effects were observed. The top row (Fig.~\ref{fig8}A1-C1) shows the extrema of the nPRCs, whereas the bottom row (Fig.~\ref{fig8}A2-C2) summarizes the extrema of the corresponding nARCs. For perturbations applied exclusively to the excitatory population (Fig.~\ref{fig8}A1 and A2), the phase response remained relatively robust over the entire range of synaptic delays. Although the magnitude of both positive and negative phase shifts gradually decreased with increasing delay, the overall balance between phase advances and delays was largely preserved. In contrast, the amplitude response exhibited a pronounced dependence on synaptic delay. Short delays produced large positive amplitude responses, indicating substantial transient enhancement of population activity, whereas increasing the delay progressively reduced the magnitude of this enhancement until only weak amplitude modulation remained for the longest delays. These findings suggest that delayed inhibition has only a modest influence on collective phase resetting induced by excitatory stimulation, while strongly regulating the resulting modulation of oscillation amplitude.

A markedly different pattern emerged when perturbations were delivered to the inhibitory population (Fig.~\ref{fig8}B1 and B2). Increasing synaptic delay substantially reduced the magnitude of the positive phase response, whereas the negative phase component became progressively smaller after an initially steep decrease, reflecting the delay-dependent reshaping of the population phase response. More strikingly, inhibitory stimulation consistently produced negative amplitude responses across all delays, indicating transient suppression of population oscillations. The strongest amplitude reduction occurred at intermediate delays ($d_{\rm syn}\approx3.0$-$4.0~\rm ms$), after which the magnitude of suppression gradually decreased. These observations demonstrate that inhibitory perturbations predominantly act to attenuate collective activity and that the effectiveness of this suppression critically depends on the timing of inhibitory feedback.

When the entire excitatory-inhibitory network was perturbed simultaneously (Fig.~\ref{fig8}C1 and C2), the phase response combined characteristics of both excitatory and inhibitory stimulation. The maximum phase advance was largest for the shortest synaptic delay and gradually decreased as the delay increased, whereas the negative phase component remained comparatively small throughout the investigated range. The amplitude response likewise showed the strongest enhancement for short delays, followed by a monotonic reduction with increasing delay. Compared with excitatory stimulation alone, however, whole-network stimulation generated substantially larger positive amplitude responses, indicating that the combined excitation of both neuronal populations amplifies the transient collective response despite the concurrent recruitment of inhibition.

Taken together, Fig.~\ref{fig8} provides a concise summary of the distinct roles of excitatory and inhibitory perturbations in shaping collective network dynamics. Excitatory stimulation primarily enhances oscillation amplitude while producing relatively stable phase resetting across delays. In contrast, inhibitory stimulation predominantly suppresses oscillation amplitude and exhibits stronger delay-dependent modulation of the phase response. Whole-network stimulation combines these effects, producing robust phase advances together with pronounced amplitude enhancement at short delays that progressively weakens as delayed inhibitory feedback becomes more dominant. Overall, these results identify synaptic delay as a key parameter governing the balance between phase resetting and amplitude modulation in PING networks, thereby regulating the sensitivity of collective oscillations to transient external inputs.


\section{Discussion}

In this study, we investigated how synaptic delays shape the collective response of oscillatory excitatory-inhibitory spiking networks to transient perturbations. By systematically varying the synaptic delay within a PING network and quantifying both nPRCs and nARCs, we demonstrated that synaptic delay regulates not only the intrinsic dynamics of the excitatory-inhibitory network but also its susceptibility to external inputs. Specifically, increasing synaptic delay slowed gamma oscillations while enhancing population synchrony, revealing a trade-off between oscillation frequency and coherence. More importantly, the response to external perturbations depended strongly on both the timing of stimulation within the oscillation cycle and the neuronal population receiving the perturbation. These findings identify synaptic delay as a fundamental determinant of collective phase resetting and amplitude modulation in cortical-like oscillatory networks.

One of the central observations of this work is that excitatory and inhibitory perturbations influence network dynamics through distinct mechanisms. Perturbations delivered to the excitatory population produced relatively robust phase responses across the investigated range of synaptic delays, whereas their influence on oscillation amplitude depended strongly on delay. This result suggests that transient excitation primarily advances or delays the ongoing rhythm by directly recruiting excitatory neurons, while the delayed inhibitory feedback largely preserves the qualitative structure of collective phase resetting. In contrast, inhibitory perturbations produced substantially stronger delay-dependent modulation of both phase and amplitude responses. As synaptic delay increased, the collective phase response became progressively more biphasic, indicating that delayed inhibition reshapes the temporal window during which inhibitory inputs can effectively modify the oscillatory cycle. These observations emphasize that inhibition is not simply a suppressive mechanism but a dynamic regulator whose effectiveness depends critically on its timing relative to the evolving excitatory population activity~\cite{dodla2006well,hatamian2026modulation}.

The complementary information provided by the nARCs further highlights the importance of considering amplitude responses alongside phase resetting. Whereas phase responses characterize changes in oscillation timing, amplitude responses capture transient changes in the strength and coherence of population oscillations. Our simulations showed that excitatory perturbations generally enhanced oscillation amplitude, whereas inhibitory perturbations predominantly reduced it, with the magnitude of both effects depending strongly on synaptic delay. These findings suggest that targeting inhibitory populations may offer a more effective approach for modulating gamma rhythms, in agreement with previous findings~\cite{akao2018relationship}. Moreover, the extrema of the nARCs closely paralleled the delay-dependent changes in network synchrony observed under baseline conditions. This relationship suggests that nARCs provide a macroscopic measure of the network's collective susceptibility, linking transient perturbation responses to the underlying synchronization state. The summary analysis presented in Fig.~\ref{fig8} further demonstrates that synaptic delay systematically regulates the balance between phase resetting and amplitude modulation across all stimulation paradigms, providing an integrated description of the network's dynamical sensitivity.

Our results also provide insight into the dynamical role of synaptic delays themselves. In the present model, increasing synaptic delay prolongs the interval during which recurrent excitation can accumulate before inhibitory feedback arrives. Consequently, excitatory neurons fire more coherently before the delayed inhibitory reset, leading to slower but more synchronized oscillations. This same mechanism explains the delay dependence of the perturbation responses. Because external stimulation interacts with the excitatory-inhibitory feedback loop at different stages of the oscillatory cycle, altering the delay changes the temporal overlap between the perturbation and the arriving inhibitory feedback. Synaptic delay therefore regulates not only the intrinsic oscillation period but also the network's effective temporal integration window for external inputs. These findings complement previous theoretical studies~\cite{ernst1995synchronization,crook1997role,brunel2000dynamics,roxin2005role} demonstrating the role of delays in synchronization and oscillation generation by extending their influence to collective phase and amplitude sensitivity.

From a broader perspective, these findings have implications for understanding how cortical circuits respond to transient inputs, such as sensory stimuli or external stimulations. Delayed interactions may allow networks to selectively amplify or suppress oscillations depending on the timing of inputs, providing a mechanism for flexible control of synchronization. This is particularly relevant in the context of brain stimulation protocols seeking to modulate pathological or physiological neuronal oscillations, where timing relative to ongoing oscillations is known to be critical~\cite{krause2022brain,taghavi2025tuning}. For example, non-invasive and invasive stimulation techniques, including transcranial alternating current stimulation (tACS), and deep brain stimulation (DBS), increasingly exploit the phase dependence of neuronal oscillations to maximize therapeutic efficacy~\cite{cagnan2017stimulating,holt2019phase,duchet2020phase,west2022stimulating,krause2022brain}. Our results suggest that the effectiveness of such stimulation protocols may depend not only on the phase of stimulation but also on the underlying time delays that determine network responsiveness~\cite{rosenblum2004delayed,popovych2005effective,popovych2017pulsatile,madadi2023decoupling,de2024disrupting}. In particular, the pronounced differences between excitatory and inhibitory perturbations indicate that stimulation targeting distinct neuronal populations - or preferentially engaging excitatory versus inhibitory circuits - may produce qualitatively different effects on oscillation timing and synchronization. Moreover, the strong relationship between synaptic delay and amplitude modulation suggests that individual variability in inhibitory kinetics or conduction delays could contribute to the heterogeneous responses frequently observed in clinical brain stimulation studies. Incorporating delay-dependent network sensitivity into computational models may therefore improve stimulation strategies by enabling protocols that account for both oscillatory phase and intrinsic circuit timing.

Although the model captures essential features of PING-mediated gamma oscillations, several limitations should be acknowledged. First, the network consists of homogeneous LIF neurons with identical synaptic parameters within each neuronal population, whereas cortical circuits exhibit substantial heterogeneity in intrinsic electrophysiological properties, connectivity, and synaptic kinetics~\cite{dahmen2026heterogeneity,dalla2026spatially}. Second, all synapses share a single fixed delay, while real cortical networks contain broad distributions of axonal and dendritic conduction delays that depend on cell type, distance, and myelination. Third, the model includes only one inhibitory interneuron population and therefore does not account for the diverse inhibitory subclasses known to shape cortical oscillations through distinct temporal dynamics~\cite{whittington2003interneuron,keeley2017modeling,cardin2018inhibitory}. Finally, external perturbations were modeled as brief Gaussian pulse packets delivered under stationary background input. More realistic stimulation paradigms may involve temporally structured, noisy, or adaptive inputs that could interact with ongoing oscillations in more complex ways. While these simplifications facilitate mechanistic interpretation, they necessarily limit the direct quantitative translation of our results to biological circuits.

Several directions for future research naturally follow from this work. Introducing heterogeneous or distributed synaptic delays would allow investigation of how delay variability influences collective phase sensitivity and oscillation stability. Incorporating synaptic plasticity, particularly spike-timing-dependent plasticity (STDP), could reveal how long-term network reorganization modifies phase and amplitude responses over extended timescales~\cite{tass2006long,kromer2020long,madadi2023decoupling,madadi2025rhythmic}. Extending the model to include multiple inhibitory interneuron populations with distinct synaptic kinetics would enable examination of how interacting inhibitory pathways shape collective oscillations across multiple frequency bands. Finally, combining the present framework with realistic models of brain stimulation may provide a useful theoretical basis for designing adaptive stimulation protocols that exploit delay-dependent network responsiveness to selectively enhance or suppress pathological synchronization.

Overall, our study demonstrates that synaptic delay serves as a fundamental control parameter governing the responsiveness of oscillatory excitatory-inhibitory networks. Beyond determining oscillation frequency and synchronization, synaptic delay regulates how transient perturbations reshape the timing and coherence of collective activity. By integrating phase resetting, amplitude modulation, and delay-dependent synchronization within a unified computational framework, this work provides new insight into the dynamical principles that govern cortical oscillations and offers a foundation for understanding and controlling rhythmic brain activity through targeted stimulation.


\section*{CRediT Author Statement}

\textbf{Parsa Shahab Rad:} Methodology, Formal analysis, Visualization, Investigation, Writing - review \& editing. \textbf{Mojtaba Madadi Asl:} Methodology, Formal analysis, Visualization, Writing - original draft, Writing - review \& editing, Project administration. \textbf{Alireza Valizadeh:} Conceptualization, Methodology, Formal analysis, Writing - original draft, Writing - review \& editing, Supervision.


\section*{Declaration of Competing Interests}

The authors declare that the research was conducted in the absence of any commercial or financial relationships that could be construed as a potential conflict of interest.



\section*{Funding}

No funding was received for conducting this study.


\section*{Data Availability}

All data used to produce the figures were generated via numerical simulations. The simulation code is publicly accessible at \href{https://github.com/parsa-shr/Synaptic-delay-phase-amplitude-response}{https://github.com/parsa-shr/Synaptic-delay-phase-amplitude-response}.



{\footnotesize\bibliography{references}}
\bibliographystyle{vancouver}
\addcontentsline{toc}{section}{References}

\end{document}